\documentstyle[aps,preprint]{revtex}
\begin{document}
\tightenlines
\draft
\title{Fresh inflation and decoherence of super Hubble fluctuations}
\author{Mauricio Bellini\footnote{E-mail address: mbellini@mdp.edu.ar;
bellini@ginette.ifm.umich.mx}}
\address{Instituto de F\'{\i}sica y Matem\'aticas, \\
Universidad Michoacana de San Nicol\'as de Hidalgo, \\
AP:2-82, (58041) Morelia, Michoac\'an, M\'exico.}

\maketitle
\begin{abstract}
I study a stochastic approach to the recently introduced fresh inflation
model for super Hubble scales. 
I find that the state
loses its coherence at the end of the fresh inflationary period as a 
consequence of the damping of the interference function in the reduced
density matrix.
This fact should be a consequence of (a) the relative evolution of
both the scale factor and the horizon and (b)
the additional thermal and dissipative effects.
This implies a relevant difference with respect to supercooled inflationary
scenarios which require a very rapid expansion of the scale factor
to the give decoherence of super Hubble fluctuations.
\end{abstract}
\vskip 2cm
\noindent
{\rm Pacs:} 98.80.Cq \\
\vskip 2cm
{\em Introduction and review of fresh inflation:}
Most of inflationary models are 
based on the dynamics of a quantum field
undergoing a phase transition\cite{a}. The exponential expansion of the
scale parameter gives a scale-invariant spectrum naturally. This is one
of the many attractive features of the inflationary universe, particulary
with regard to the galaxy formation problem. It arises from the
fluctuations of the inflaton, the quantum field that induces inflation.
This field can be semiclassicaly expanded in terms of its expectation value
plus another field, which describes the quantum fluctuations. The standard
slow-roll inflation model separates expansion and reheating into two
distinguished time periods. It is first assumed that exponential expansion
from inflation places the universe in a supercooled phase. Subsequently
the universe in reheated. Two outcomes arise from such a
scenario. First, the required density perturbations in this cold
universe are left to be created by the quantum fluctuations of the
inflaton. Second, the temperature cliff after expansion requires a 
temporally localized mechanism that rapidly sufficient
distributes 
vacuum energy for reheating. So, the scalar field oscillates near
the minimum of its effective potential and produces elementary particles.
This process is completed when all the energy of the classical scalar
field transfers to the thermal energy of elementary particles.

On the other hand, warm inflation takes into account separately the matter
and radiation energy densities that are responsible for the fluctuations
of temperature. In this scenario, the inflaton field interacts with other
particles that are in a thermal bath with a mean temperature smaller
than the grand unified theory (GUT) critical temperature\cite{1,2}. 
The problem with warm inflation is that, at
the beginning of the universe, the thermal bath is unjustifiedly introduced
in the framework of chaotic initial conditions needed to give a natural
beginning to the universe. In this sense, chaotic inflation\cite{3}
provides a more successful and natural mechanism to describe the
initial conditions in the universe. In the warm inflation scenario,
slow-roll conditions are induced through dissipative damping with no
requirement of an ultraflat potential. Furthermore, this scenario
differs from standard inflation
in that reheating is no longer required.
In warm inflation, exponential expansion and thermal energy production
occur together.

Very recently, a new model of inflation called {\em fresh inflation}
was proposed\cite{4}. 
Fresh inflation can be viewed as a
``unification'' of both chaotic and warm inflation scenarios.
Fresh inflation incorporates the following characteristics of chaotic
and warm inflationary scenarios:
\begin{itemize}
\item As in chaotic inflation, the universe begins from an unstable
primordial matter field perturbation with energy density nearly
$M^4_p$ ($M_p =1.2 \times GeV$  is the Planckian mass) and chaotic
initial conditions. Furthermore, initially the universe is not thermalized
so that the radiation energy density when inflation starts is zero 
[$\rho_r(t=t_0)=0$]. We understand the initial time to be
the Planckian time
$G^{1/2}$.
Later, the universe will describe a second-order
phase transition. In other words, the inflaton rolls down towards the
minimum of the potential. 
\item Particle production and heating occur together during the rapid
expansion of the universe, so that the radiation energy density grows
during fresh inflation ($\dot\rho_r >0$).
The interaction between the inflation field and the particles produced
during inflation provides slow-rolling of the inflaton field.
So, in the fresh inflationary model (as in warm inflation), the
slow-roll conditions are physically well justified.
\item The decay width of the produced particles grows with time. When
the inflaton approaches the minimum of the potential, there is
no oscillation of the inflaton 
around the minimum energetic configuration due to dissipation being
too large ($\Gamma \gg H$) at the end of fresh inflation. 
Hence, the reheating
period avoids fresh inflation (as in warm inflation).
\end{itemize}

In this work, I study decoherence of
the state that describes super
Hubble matter field fluctuations during fresh inflation. 
Decoherence of super Hubble fluctuations has been studied in the framework
of supercooled inflation\cite{sta}.
It is well known that the evolution
of the redefined coarse-grained field is described by a 
second-order stochastic
equation. 
This topic was studied in \cite{5} for supercooled
inflation and also in \cite{6} for warm inflation.
The effective Hamiltonian related to this stochastic
equation can be described in a such away that
the Schr\"odinger equation for the system can be written.
The wave function that describes this system is $\Psi(\chi_{cg},t)$, where
$\chi_{cg}$ denotes the coordinate 
(i.e., the redefined coarse-grained field fluctuations).
The effect of summing over unobservable degrees of freedom (ultraviolet
sector) of small inhomogeneous modes reduces to a multiplication of the
reduced density matrix by an interference term of the form $
e^{-\alpha(t) [\chi_{cg}-\chi'_{cg}]^2}$, where $\alpha(t)$ is a
time-dependent function and $(\chi_{cg}, \chi'_{cg})$ are two different
configurations of the redefined coarse-grained field fluctuations.
If the interaction of $\chi_{cg}$ with the environment $\chi_s$ ($\chi_s$ takes
into account only the short modes) is sufficiently strong to damp quantum
interference, the system decoheres.
The time evolution of the decoherence function as a consequence
of damping of the interference function is the main feature
to be analyzed in this paper.

I consider a Lagrangian for a $\phi$-scalar field minimally coupled
to gravity, which also interacts with another $\psi$-scalar field
by means of a Yukawa interaction,
\begin{equation}\label{1}
{\cal L} = - \sqrt{-g} \left[\frac{R}{16\pi G} +\frac{1}{2}
g^{\mu\nu} \phi_{,\mu}\phi_{,\nu} + V(\phi)\right] +
{\cal L}_{int},
\end{equation}
where $g^{\mu\nu}$ is the metric
tensor, $g$ is its determinant and $R$ is the scalar curvature. 
The interaction Lagrangian is given by 
${\cal L}_{int} \sim -g^2 \phi^2 \psi^2$
Furthermore,
the indices $\mu,\nu$ take the values $0,..,3$
and the gravitational constant is $G=M^{-2}_p$ (where
$M_p = 1.2 \times 10^{19} \  GeV$ is the Planckian mass).
The Einstein equations for a globally flat, isotropic, and homogeneous
universe described by a Friedmann-Robertson-Walker metric
$ds^2 = -dt^2 + a^2(t) dr^2$ are given by
\begin{eqnarray}
3 H^2 & =& 8\pi G\left[ \frac{\dot\phi^2}{2} + V(\phi) +
\rho_r\right], \label{4} \\
3H^2 + 2 \dot H & = & -8\pi G\left[\frac{\dot\phi^2}{2}-
V(\phi) + \rho_r\right], \label{5}
\end{eqnarray}
where $H={\dot a\over a}$ is the Hubble parameter and $a$ is the
scale factor of the universe. 
The overdot denotes
the derivative with respect to the time.
On the other hand, if $\delta=\dot\rho_r+4H\rho_r$
describes the interaction between the inflaton and the bath,
the equations of motion for $\phi$ and $\rho_r$
are
\begin{eqnarray}
&& \ddot\phi+3H\dot\phi+V'(\phi) + \frac{\delta}{\dot\phi}=0,\label{6}\\
&& \dot\rho_r+4H\rho_r-\delta=0. \label{7}
\end{eqnarray}
As in a previous paper\cite{4}, I will consider a Yukawa interaction
$\delta = \Gamma(\theta) \  \dot\phi^2$, where $\Gamma(\theta)=
{g^4_{eff}\over 192\pi}\theta$\cite{7} and $\theta \sim \rho^{1/4}_r$ 
is the temperature of the bath.
Slow-roll conditions must be imposed to assure nearly de Sitter solutions
for an amount of time, long enough to solve the 
flatness and horizon problems. If $p_t={\dot\phi^2 \over 2}+{\rho_r\over 3} - V(\phi)$
is the total pressure and $\rho_t=\rho_r+{\dot\phi^2 \over 2}+V(\phi)$ is the
total energy density, the parameter $F={p_t+\rho_t \over \rho_t}$
which describes the evolution of the universe during inflation\cite{8} is
\begin{equation}\label{8}
F= - \frac{2\dot H}{3 H^2} = \frac{\dot\phi^2+\frac{4}{3} \rho_r}{
\rho_r+ \frac{\dot\phi^2}{2}+V}.
\end{equation}
When fresh inflation starts (at $t=G^{1/2}$), the radiation energy 
density is zero, so that $F\ll 1$. 

In this paper, I will consider the parameter $F$ as a constant.
From the
two equalities in eq. (\ref{8}), one obtains the following equations:
\begin{eqnarray}
&& \dot\phi^2 \left(1-\frac{F}{2}\right) +
\rho_r\left(\frac{4}{3}-F\right)-F \  V(\phi)=0, \label{9} \\
&& H= \frac{2}{3 \int F \  dt}.\label{10}
\end{eqnarray}
Furthermore, because of $\dot H = H'\dot\phi$ (here the prime denotes
the derivative with respect to the field), from the first equality
in eq. (\ref{8}) it is possible to obtain the equation that
describes the evolution for $\phi$,
\begin{equation}\label{11}
\dot\phi=-\frac{3 H^2}{2 H'}F,
\end{equation}
and replacing eq. (\ref{11}) in eq. (\ref{9}), the radiation energy
density can be described as functions of $V$, $H$ and $F$\cite{4}
\begin{equation}\label{12}
\rho_r = \left(\frac{3F}{4-3F}\right) V - \frac{27}{8}
\left(\frac{H^2}{H'}\right)^2 \frac{F^2(2-F)}{(4-3F)}.
\end{equation}
Finally, replacing eqs. (\ref{11}) and (\ref{12}) in eq. (\ref{4}),
the potential can be written as
\begin{equation}\label{13}
V(\phi) = \frac{3}{8\pi G} \left[\left(\frac{4-3F}{4}\right)
H^2 + \frac{3\pi G}{2} F^2\left(\frac{H^2}{H'}
\right)^2\right].
\end{equation}

Fresh inflation was proposed for a global group $O(n)$ involving
a single $n$-vector multiplet of scalar fields $\phi_i$\cite{9},
such that making $(\phi_i\phi_i)^{1/2}\equiv \phi$, the effective
potential $V_{eff}(\phi,\theta)=V(\phi)+\rho_r(\phi,\theta)$ can be
written as
\begin{equation}\label{14}
V_{eff}(\phi,\theta) = \frac{{\cal M}^2(\theta)}{2} \phi^2+
\frac{\lambda^2}{4}\phi^4,
\end{equation}
where ${\cal M}^2(\theta) = {\cal M}^2(0)+{(n+2) \over 12} \lambda^2\theta^2$
and $V(\phi)= {{\cal M}^2(0)\over 2} \phi^2+{\lambda^2 \over 4}\phi^4$.
Furthermore, ${\cal M}^2(0) >0$ is the squared mass at
zero temperature, which is given by ${\cal M}^2_0$ plus renormalization
counterterms in the potential ${1 \over 2} {\cal M}^2_0 (\phi_i\phi_i)+
{1 \over 4} \lambda^2 (\phi_i\phi_i)^2$\cite{9}.
I will take 
into account the case without symmetry breaking, ${\cal M}^2(\theta) >0$
for any temperature $\theta$, so that
the excitation spectrum consists of $n$ bosons with
mass ${\cal M}(\theta)$.
Note that the effective potential (\ref{14}) 
is invariant under $\phi \rightarrow -\phi$ 
reflections and $n$ is the number of created particles due to
the interaction of $\phi$ with the particles in the thermal bath, such
that\cite{4}
\begin{equation}\label{15}
(n+2) = \frac{2\pi^2}{5\lambda^2}g_{eff} \frac{\theta^2}{\phi^2},
\end{equation}
because the radiation energy density is given by $\rho_r={\pi^2 \over 30}
g_{eff} \theta^4$  ($g_{eff}$ denotes the effective degrees
of freedom of the particles and it is assumed that $\psi$ has no                                                
self-interaction). A particular solution of eq. (\ref{13}) is
\begin{equation}
H(\phi) =4 \sqrt{\frac{\pi G}{3(4-3F)}} \  {\cal M}(0)  \  \phi.\label{16} \\
\end{equation}
From eq. (\ref{10}), and due to 
$H=\dot a/a$, one obtains the scale factor
as a function of time
\begin{equation}\label{18}
a(t) \sim  t^{\frac{2}{3F}}.
\end{equation}
Furthermore, the number of e-folds during fresh inflation is
\begin{equation}\label{19}
N(t) = \left.\frac{2}{3F} {\rm ln}(t)\right|^{t_e}_{t_s},
\end{equation}
which, due to the fact that $\phi=\lambda^{-1} t^{-1}$, can be
written as a function of $\phi$: $N(\phi) = \left.{2 \over 3 F}
{\rm ln}\left({1\over \lambda\phi}\right)\right|^{\phi_e}_{\phi_s}$,
which grows as $\phi$ decreases. 
Here, $(t_s, t_e)$ are the starting and ending values of time
and $(\phi_s, \phi_e)$ are the starting and ending values of $\phi$
(for $t_e > t_s$ and $\phi_e < \phi_s$).
Taking $g_{eff} \simeq 10^2$ and
$\phi_e \simeq 10^{-4} G^{-1/2}$ one obtains the number of
created particles at the end of fresh inflation\cite{4}:
$n_e \simeq 4 \times 10^8$.

{\em Coarse-graining, interference and decoherence of the fluctuations:}
Classical physics is characterized by the fact that one can assign a
probability to each possible history of the system. In contrast, in quantum
physics one must assign a complex probability amplitude to configuration
variables, because there are no trajectories in quantum theory.
When combined with the superposition principle, this implies the existence
of quantum interference effects. On the other hand, it is an empirical fact
that these effects are not seen at a macroscopic (or classical) level.
One of the most popular ways of fixing this problem is by means of 
decoherence arguments, whose essence is the following:
One can consider that the original system is part of a more complicated
world and interacts with an ``environment'' formed by unobserved
(or irrelevant) degrees of freedom. Then, under some circumstances, it is
possible to show that the quantum interference effects on the system
can be suppressed by
the interaction with the environment. The system thus decoheres (it 
loses quantum coherence) and can be described as a statistical mixture of 
noninterfering branches\cite{paz}. Other possible descriptions involve
the Schr\"odinger\cite{hami,mijic} or Wigner\cite{habib} pictures.

The equation of motion for
the matter field fluctuations $\delta\phi(\vec x,t)$ is
\begin{equation}
\ddot{\delta\phi} + \left(3 H+\Gamma\right)\dot{\delta\phi}-
\frac{1}{a^2} \nabla^2\delta\phi + V''(\phi) \delta\phi=0,
\end{equation}
where $V''$ denotes the second derivative of the potential with respect to
the field. This equation can be simplified by means of the
map $\chi = e^{3/2 \int(H+\Gamma/3) \  dt} \delta\phi$ 
\begin{equation}
\ddot\chi_k + \omega^2_k(t) \  \chi_k =0,
\end{equation}
where I supposed a Fourier expansion for the scalar field $\chi$
in term of its modes $\chi_k = e^{i \vec k.\vec x} \xi_k(t)$. Furthermore,
$\omega^2_k(t)=a^{-2} \left[k^2 - k^2_0(t)\right]$ is the squared
frequency for each mode with wave number $k$, and $k_0(t)$
gives the time-dependent wave number that separates the infrared (IR)
and ultraviolet (UV) sectors. The IR sector describes the 
large-scale fluctuations (or super Hubble fluctuations)
and the UV sector 
takes into account the matter field fluctuations on
sub Hubble scales ($k^2 \gg k^2_0$). The squared time-dependent
wave number $k^2_0(t)$ is given by the expression
\begin{equation}\label{199}
k^2_0 = a^2\left[\frac{9}{4}
\left(H+\frac{\Gamma}{3}\right)^2 + 3 \left(\dot H + \frac{\dot\Gamma}{3}
\right) - V''\right],
\end{equation}
where $\Gamma={g^4_{eff} \over 192\pi} \theta$ and the temperature
$\theta(t)$ is given by
\begin{equation}
\theta(t) = \frac{192\pi}{g^4_{eff} \lambda^2} \left\{ {\cal M}^2(0)
\lambda^2 t + t^{-1} \left[ \lambda^2 \frac{(9 F^2 - 18 F +8)}{
(4-3F)^2}+ {\cal M}^2(0) \pi G \frac{(192 F^2 - 72 F^3 -96)}{
(4-3F)^2}\right]\right\},
\end{equation}
which, for late times, increases linearly with time.
Note that $\theta(t)$ becomes zero at $t=G^{1/2}$ (we use Planckian units).

The redefined fluctuations $\chi$ can be decomposed in long-and
short-wavelength components $\chi_{cg}$ and $\chi_s$, respectively.
In our case, we are interested in the study of decoherence for
super Hubble (cosmological) scales. So, the relevant degrees of
freedom will be $k^2 \ll k^2_0$, 
which are incorporated in the
coarse-grained field $\chi_{cg}$. The field $\chi_s$ takes into account
the unobserved degrees of freedom (environment).
The redefined
coarse-grained field, can be written as a Fourier expansion in the $k$ space,
\begin{equation}
\chi_{cg}(\vec x,t) 
= \frac{1}{(2\pi)^{3/2}} {\Large\int} d^3k \  \theta(\epsilon k_0-
k) \left[a_k \chi_k(\vec x,t) + a^{\dagger}_k \chi^*_k(\vec x,t)\right],
\end{equation}
where $\epsilon \ll 1$ is a dimensionless constant and $\chi_{k}(\vec x,t)$
are the modes of $\chi_{cg}$. Furthermore, the annihilation and creation
operators $(a_k,a^{\dagger}_k)$, satisfy the commutations relations
$\left[a_k,a^{\dagger}_{k'}\right]=\delta^{(3)}(k-k')$ and
$\left[a_k,a_{k'}\right]=\left[a^{\dagger}_k,a^{\dagger}_{k'}\right]=0$.

The stochastic equation for the redefined coarse-grained field,
$\ddot\chi_{cg}+ \omega^2_k(t) \chi_{cg} + \xi_c(\vec x,t)=0$, can be
approximed to the zero-mode stochastic equation for very large
scale fluctuations (i.e., for $k^2\ll k^2_0$),
\begin{equation}\label{29}
\ddot\chi_{cg} - \mu^2(t) \chi_{cg} + \xi_c(\vec x,t)=0,
\end{equation}
where $\xi_c = -\epsilon \left[{d\over dt} \left(\dot k_0 \eta\right)+
2 \dot k_0 \kappa\right]$ is the effective noise and
($\eta$, $\kappa$) are given by
\begin{eqnarray}
\eta(\vec x,t) & =& \frac{1}{(2\pi)^{3/2}} {\Large\int} d^3k \  \delta(\epsilon k_0-
k) \  \left[a_k \chi_k + h.c.\right], \\
\kappa(\vec x,t) & =& \frac{1}{(2\pi)^{3/2}} {\Large\int} d^3k \  \delta(\epsilon k_0-
k) \  \left[a_k \dot\chi_k + h.c.\right].
\end{eqnarray}
This noise arises from 
(a) the inflow of short-wavelength modes produced
by the relative evolution of both the horizon and the scale factor of
the universe, and (b) the created particles during fresh inflation, which are
thermalized.

The effective Hamiltonian for eq. (\ref{29}) is
\begin{equation}\label{31}
H_{eff}(\chi_{cg},t) = \frac{1}{2} P^2_{cg} -\mu^2(t) \chi^2_{cg} +
\xi_c \chi_{cg},
\end{equation}
where $P_{cg}\equiv \dot\chi_{cg}$. 
It is here that we depart from the Heisenberg to the Schr\"odinger
picture. The derivative operator representation of the momentum
ensures that the Heisenberg and the Schr\"odinger pictures reflect
the same physics. Furthermore, $\xi_c$ is an external classical force in
the quantum Hamiltonian (\ref{31}).
With this assumption 
we can write the Schr\"odinger
equation\cite{mijic}
\begin{equation}\label{32}
i \frac{\partial}{\partial t} \Psi(\chi_{cg},\xi_c,t) =
-\frac{1}{2} \frac{\partial^2}{\partial\chi^2_{cg}} \Psi(\chi_{cg},\xi_c,t) +
\left[ -\frac{\mu^2(t)}{2} \chi^2_{cg}
+ \xi_c \chi_{cg}\right] \Psi(\chi_{cg},\xi_c,t).
\end{equation}
Here, $\Psi(\chi_{cg},\xi_c,t)$ is the wave
function that describes the system in the $\chi_{cg}$ representation.
Notice that the states $\Psi(\chi_{cg},\xi_c,t)$ are fiducial because no pure
states can be attributed to a quantum subsystem that is entangled with other
subsystems.
A method to solve this equation can be found in \cite{010}.
If the initial state is taken to be a harmonic-oscillator ground
state and the external (stochastic) force $\xi_c$ is set to zero,
then, because
of the time-dependent frequency ($\omega^2_{k=0} = -\mu^2$), the
solutions of the Schr\"odinger equation are squeezed states at zero
momentum. In our case, the state becomes highly squeezed due to the accelerated
expansion of the universe and the $\phi-\psi$ Yukawa interaction, which
introduces additional friction terms (proportional to $\Gamma$, $\Gamma^2$,
and $\dot\Gamma$ or $\theta$, $\theta^2$, and $\dot\theta$) in $k^2_0(t)$
[see eq. (\ref{199})].
On the other hand, if the frequency is time-independent but the external
force is nonzero, the solutions are coherent states\cite{011}. As can
be seen readily, equation (\ref{32}) combines both situations.

The solution for the Schr\"odinger equation (\ref{32}) is given by\cite{10}
\begin{eqnarray}
\Psi(\chi_{cg},\xi_c,t) &=& \frac{1}{(2\pi)^{1/4}\Delta^{1/2}}
e^{-\frac{1}{4\Delta^2}\left[\chi_{cg}-\chi_{cl}\right]^2}
e^{i\frac{\chi^2_{cg}}{4\Delta^2} \left[2\frac{\dot B}{B}\Delta^2 +
\frac{{\cal R}(t)}{\sigma^2}\right]} \nonumber \\
& \times & e^{i \frac{\chi_{cg}}{\Delta^2} \left[\Delta^2(P_{cl} -
\frac{\dot B}{B} \chi_{cl}) - \frac{{\cal R}(t)\chi_{cl}}{2\sigma^2}\right]}
e^{i \gamma(t)},
\end{eqnarray}
where $\gamma(t)$ is an arbitrary phase and, in our case, the
time-dependent functions $B$
are given by the zero modes $\xi_0$ of the redefined
coarse-grained fluctuations $\chi_{cg}$ 
[i.e., for $B(t) = \xi_{k=0}(t)\equiv \xi_0(t)$]\cite{10}:
\begin{eqnarray}
{\cal R}(t) & =& {\Large\int}^t dt' \frac{1}{2 \xi^2_0(t')}, \label{a} \\
\Delta^2(t) & = & \frac{\xi^2_0(t)}{\sigma^2} \left[\sigma^4 +
{\cal R}^2(t)\right], \label{b}
\end{eqnarray}
where $\sigma^2 = \Delta^2(t_0)$ is a constant
and $\Delta^2$ gives the squared dispersion. 
The reduced density matrix related to the wave function $\Psi(\chi_{cg},
\xi_c,t)$ for two different configurations of $\chi_{cg}$ and
$\chi'_{cg}$ is\cite{paz}
\begin{equation}
\rho_{red}\left(\chi_{cg},\chi'_{cg},t\right) =
{\Large \int} d\xi_c \  \rho_{red}\left(\chi_{cg},\chi'_{cg},\xi_c,t\right),
\end{equation}
where
$\rho_{red}\left(\chi_{cg},\chi'_{cg},\xi_c,t\right)
=\Psi^*\left(\chi_{cg},\xi_c,t\right)
\Psi\left(\chi'_{cg},\xi_c,t\right)$ such that
\begin{equation}\label{inter}
\rho_{red}\left(\chi_{cg},\chi'_{cg},t\right) =
{\Large \int} d\xi_c \  \rho_0\left(\chi_{cg},\chi'_{cg},\xi_c,
t\right) \  {\cal I}\left(\chi_{cg},\chi'_{cg},t\right),
\end{equation}
and
\begin{eqnarray}
&& \rho_0\left(\chi_{cg},\chi'_{cg},\xi_c,t\right)= \frac{1}{(2\pi)^{1/2}
\Delta} e^{-\frac{1}{4 \Delta^2} \left[\left(\chi_{cg}-\chi_{cl}\right)^2+
\left(\chi'_{cg}-\chi_{cl}\right)^2\right]} \nonumber \\
&\times & e^{-i\frac{1}{4 \Delta^2} \left[\Delta^2\left( P_{cl}-
\frac{\dot\xi_0}{\xi_0} \chi_{cl}\right)-\frac{{\cal R}(t)\chi_{cl}}{
2\sigma^2}\right]\left[\chi_{cg}-\chi'_{cg}\right]} \nonumber \\
& \times & e^{-i\frac{1}{\Delta} \left[\frac{\dot\xi_0}{\xi_0} \Delta+
\frac{{\cal R}(t)}{2 \sigma^2 \Delta}\right] \chi_{cg} \chi'_{cg}}.
\end{eqnarray}
The interference function\cite{habib} that 
multiplies $\rho_0$ in eq. (\ref{inter}) is
\begin{equation}
{\cal I}\left(\chi_{cg},\chi'_{cg},t\right) =
e^{-i\frac{\sqrt{{\cal D}}}{\sqrt{2} \Delta}
\left[\chi_{cg}-\chi'_{cg}\right]^2}.
\end{equation}
where ${\cal D}(t)$ is the function
that describes decoherence
in the quantum state $\Psi$\cite{hami,10},
\begin{equation}\label{c}
{\cal D}(t) = \frac{1}{2} \left( \frac{\dot\xi_0}{\xi_0} \Delta(t) +
\frac{{\cal R}}{2\Delta \sigma^2}\right)^2.
\end{equation}
(Note that in paper \cite{habib} the authors named this function the
``decoherence
function,'' but it describes the evolution of quantum interference.)
The squared fluctuations $\left<\chi^2_{cg}\right>$ and
$\left< P_{cg}\right>$ are
\begin{eqnarray}
\left<\chi^2_{cg}\right> & = & \chi^2_{cg}(t) + \Delta^2(t), \label{alfa} \\
\left<P^2_{cg}\right> & = & P^2_{cl} + \frac{1}{4 \Delta^2(t)} + 
2 {\cal D}(t),\label{beta}
\end{eqnarray}
where $\left<P_{cg}\right> = P_{cl}$ and
$\left<\chi_{cg}\right> = \chi_{cl}$. Notice we are dealing
with the $\chi_{cg}$ representation, so that $P_{cg} \equiv
- i {\partial \over \partial \chi_{cg}}$. The second terms
in eqs. (\ref{alfa}) and (\ref{beta}) represent the quantum
fluctuations which depend only on $\mu(t)$. The third
term in eq. (\ref{beta}) represents quantum fluctuations related
to decoherence. In the case of a pure quantum
state (i.e., in a coherent state) the function ${\cal D}(t)$ becomes
zero.

All we need to find the evolution of the decoherence
function (\ref{c}) in fresh inflationary cosmology 
is to solve the equation for the zero mode function
\begin{equation}\label{aa}
\ddot\xi_0(t) - \mu^2(t) \xi_0(t)=0,
\end{equation}
and later calculate the functions ${\cal R}(t)$ and $\Delta(t)$.
The squared parameter of mass $\mu^2(t)={k^2_0 \over a^2}$, in
eq. (\ref{aa}) is
\begin{equation}
\mu^2(t) 
= \frac{1}{(4-3F)^2} \left[ A_1 + A_2 \  t^{2} + A_3 \  t^{-2}\right],
\end{equation}
where the constants $A_1$, $A_2$, and $A_3$ are given by
\begin{eqnarray}
&& A_1 =  9 F {\cal M}^2(0) \left(\frac{F}{2} - 1\right) +
\frac{{\cal M}^4(0) \pi G}{\lambda^2} \left(96 F^2 - 36 F^3 - 48\right)
+ {\cal M}^2(0) \left( \frac{1}{F} - 1+ \frac{1}{\lambda^2}\right) +1,\\
&& A_2 = \frac{{\cal M}^4(0)}{4}, \\
&& A_3 = \left\{ \frac{1}{(4-3F)^2} \left[ \frac{8}{F} - 9F^2 + 27 F -26 +
4 {\cal M}^2(0)\right.\right. \nonumber \\
&& + \frac{{\cal M}^2(0)\pi G}{\lambda^2} \left(72 F^2 + 96 -
264 F^2 - \frac{96}{F} + 192 F\right)  
+\frac{1}{(4-3F)^2}\left[
\frac{81 F^4}{4} - 81 F^3 + 117 F^2  \right.\nonumber \\
&&- 72 F+16+ \frac{{\cal M}^2(0)\pi G}{\lambda^2} \left[
1512 F^4 - 324 F^5 + 336 F^2 
- 2016 F^3+864 F - 384 \right. \nonumber \\
&& +\left.\left.\left.\frac{{\cal M}^2(0) \pi
G}{\lambda^2} \left(9216 F^2 \left(F^2-1\right) 
-  6912 F^5 + 1296 F^6+ 3456 F^3 + 2304 \right)\right]\right]\right]
\nonumber \\
&& -\left.3+\frac{1}{F} \left(\frac{1}{F} - 2\right)\right\}.\label{mu}
\end{eqnarray}
For late times, the squared parameter of mass can be approximated to 
\begin{equation}
\left.\mu^2(t)\right|_{t \gg G^{1/2}} \simeq A_2 \  t^{2}.
\end{equation}
The solution for the zero modes $\xi_{k=0}(t)$, in eq. (\ref{aa}) is
\begin{equation}
\xi_0(t) = c_1 \  \sqrt{t} \  {\cal I}_{\nu}\left[\frac{1}{2}\sqrt{A_2}t^2\right] 
+ c_2 \  \sqrt{t} \  {\cal K}_{\nu}\left[\frac{1}{2}\sqrt{A_2}t^2\right],
\end{equation}
where ${\cal I}_{\nu}$ and ${\cal Y}_{\nu}$ are the Bessel functions,
$\nu = 1/4$, and ($c_1$,$c_2$) are arbitrary constants. 
For very large times (i.e., for $t \gg 10^7 \  G^{1/2}$),
we can use the asymptotic expressions for the Bessel functions,
\begin{eqnarray}
{\cal I}_{\nu}[x] \simeq \frac{e^x}{\sqrt{2\pi x}}, \\
{\cal K}_{\nu}[x] \simeq \frac{e^{-x}}{\sqrt{2\pi x}},
\end{eqnarray}
so that the asymptotic zero mode $\xi_0$ can be approximated to
\begin{equation}
\left.\xi_0(t)\right|_{t \gg G^{1/2}} \propto \frac{1}{t^{1/2}{\cal M}(0)}
e^{\frac{{\cal M}^2(0)}{2} t^2}.
\end{equation}
Hence, at the end of fresh inflation the function ${\cal R}(t)$ becomes
[see eq. (\ref{a})]
\begin{equation}
{\cal R}(t) \sim e^{-{\cal M}^2(0) \  t^2},
\end{equation}
and the squared dispersion [see eq. (\ref{b})] is
\begin{equation}
\Delta^2(t) \simeq \frac{1}{t\  {\cal M}^2(0)} e^{{\cal M}^2(0) t^2}.
\end{equation}
Finally, the asymptotic evolution of the decoherence function ${\cal D}(t)$
is given by [see eq. (\ref{c})]
\begin{equation}
{\cal D}(t) \sim \  {\cal M}^2(0) \  t \  e^{{\cal M}^2(0) t^2},
\end{equation}
which grows with time.

Finally, it must be noted that the relevant field
configuration for the super Hubble inflaton fluctuations
during the
inflationary stage is 
$\delta\phi_{cg} = e^{-3/2\int [H(t)+\Gamma(t)/3] dt} \chi_{cg}$.
This means that the relevant decoherence function during inflation
is
\begin{equation}
\left.{\cal D}\left(\delta\phi_{cg},t\right)
\right|_{t \gg G^{1/2}} \simeq \left.{\cal D}(t)\right|_{t \gg G^{1/2}}
t^{-2/F} e^{-\frac{{\cal M}^2(0)}{2} t^2},
\end{equation}
where we have taken $\Gamma(t) \simeq {\cal M}^2(0) t$, which is a good
approximation for late times. Hence, the decoherence function
for $\delta\phi_{cg}$ go as
\begin{equation}\label{resul}
\left.{\cal D}\left(\delta\phi_{cg},t\right)
\right|_{t \gg G^{1/2}} \sim \  t^{\frac{2(F-1)}{F}} 
e^{\frac{{\cal M}^2(0)}{2} t^2},
\end{equation}
which also increases with time at the end of fresh inflation.
Note that the decoherence function is maximized for large values of
$F$. In particular, when $F>1$, the universe approaches 
a radiation-dominated regime ($F=4/3$). However, as was demonstrated
in\cite{4}, to ensure slow-roll conditions one requires $F < 0.55$.
Furthermore, the interference function ${\cal I}(\delta\phi_{cg},
\delta\phi'_{cg},t)$ is given by
${\cal I}(\delta\phi_{cg},\delta\phi'_{cg},t) = e^{-i{\sqrt{{\cal D}}\over 
\sqrt{2} \Delta}
[\delta\phi_{cg} -\delta\phi'_{cg}]^2}$\cite{habib}. In the fresh
inflationary model here considered gives
\begin{equation}
{\cal I}\left(\delta\phi_{cg},\delta\phi'_{cg},t\right)=
\exp\left\{-i
\frac{t^{\frac{F+2}{2F}}}{\sqrt{2} {\cal M}(0)} \  e^{\frac{{\cal M}^2(0)}{4} t^2}
\left[\delta\phi_{cg}-
\delta\phi'_{cg}\right]^2\right\},
\end{equation}
which decreases monotonically with time during the fresh inflationary
period. Hence, during fresh inflation the quantum interference
is damped very rapidly.

To summarize,
in this work I studied decoherence of super Hubble fluctuations 
in the recently introduced fresh inflationary scenario.
Decoherence occurs when there are no interference effects
between alternative histories. In the context of inflationary cosmology,
the system is the super Hubble fluctuations. This field takes into
account the relevant degrees of freedom of the system (coarse-grained field). 
The irrelevant (or unobserved)
degrees of freedom contain the modes of the ultraviolet sector (sub-Hubble
sector). During the inflationary expansion of the universe, this sector
works as an environment. Roughly speaking, one can say that the
environment is ``continuously measuring'' the $\delta\phi_{cg}$ variable and
that this continuous measurement process suppresses quantum interference.
It is a consequence of 
(a) the inflow of short-wavelength modes produced
by the relative evolution of both the horizon and the scale factor of
the universe (this effect is responsible for the squeezing of the 
state) and
(b) additional thermal and dissipative effects.
the latter are relevant to the dissipative (thermal) effects produced by the
interaction between the inflaton field and other $\psi$-fields
in the universe.
Both facts produce decoherence of the coarse-grained field
leading to the quantum-to-classical transition of the $\delta\phi_{cg}$ field
on super Hubble scales. This means that the stochastic description
for super Hubble fluctuations in fresh inflation is consistent.
This is the main difference
with supercooled inflation (see \cite{10}), in which the power of the
expansion of the scale factor 
must be very large (i.e., requires $p > 4.6$
when $a \sim t^p$) in order for the system
to lose its coherence during inflation.
Notice we are not dealing with correlations between coordinates
and momenta. In order to analyze whether a given wave function predicts
the existence of correlations between coordinates and momenta, many authors
proposed to examine the peaks of the Wigner function associated with
it\cite{ultimos}. However, this topic is not the subject of this
work.

\vskip .2cm
\noindent
I would like to acknowledge CONACYT (M\'exico) and CIC of Universidad
Michoacana for financial support in the form of a research grant.\\

\end{document}